# Tuning the Dzyaloshinskii−Moriya Interaction in Pt/Co/MgO heterostructures through MgO thickness

Anni Cao,[a, 1)] Xueying Zhang,[abc, 1)] Bert Koopmans,[d] Shouzhong Peng,[a] Yu Zhang,[ab] Zilu Wang,[a] Shaohua Yan,[a] Hongxin Yang,[e] and Weisheng Zhao[*ac]

[*] Corresponding authors

[a] Fert Beijing Institute, BDBC, School of Electronic and Information Engineering, Beihang University, Beijing, China

[b] Centre for Nanoscience and Nanotechnology, University Paris-Saclay, Orsay, France

[c] Beihang-Goertek Joint Microelectronics Institute, Qingdao Research Institute, Beihang University, Qingdao, China

[d] Department of Applied Physics, Institute for Photonic Integration, Eindhoven University of Technology, Eindhoven, The Netherlands

[e] Key Laboratory of Magnetic Materials and Devices, Ningbo Institute of Materials Technology and Engineering, Chinese Academy of Sciences, Ningbo, Zhejiang, China

ⓈSupporting Information

**ABSTRACT** The interfacial Dzyaloshinskii−Moriya interaction (DMI) in the ferromagnetic/heavy metal ultra-thin film structures, has attracted a lot of attention thanks to its capability to stabilize Néel-type domain walls (DWs) and magnetic skyrmions for the realization of non-volatile memory and logic devices. In this study, we demonstrate that magnetic properties in perpendicularly magnetized Ta/Pt/Co/MgO/Pt heterostructures, such as magnetization and DMI, can be significantly influenced by the MgO thickness. To avoid the excessive oxidation of Co, an ultrathin Mg layer is inserted to improve the quality of Co-MgO interface. By using field-driven domain wall expansion in the creep regime, we find that non-monotonic tendencies of the DMI field appear when changing the thickness of MgO. With the insertion of a monatomic Mg layer, the strength of DMI could reach a high level and saturates. We conjecture that the efficient control of DMI is a result of subtle changes of both Pt/Co and Co/MgO interfaces, which provides a method to optimize the design of ultra-thin structures achieving skyrmion electronics.

**INTRODUCTION**

In the past few years, the Dzyaloshinskii−Moriya Interaction (DMI) attracted significant interests[1] because it is one of the key ingredients to create magnetic skyrmions and chiral domain walls (DWs), which are promising for the next generation of high-speed magnetic storage devices. After more than two decades of theoretical studies[2–4], magnetic skyrmions were first observed at low temperature in hexagonal lattices in non-centrosymmetric crystals[5–7] and magnetic multilayers[8–10]. On the other hand, interfacial DMI induced by symmetry breaking at the interface appears to be particularly important. In addition, DW motion closed to 1000 m/s has been observed[11] The polarized spin current from the heavy metal layer due to the spin Hall effect combined with a Néel-type DW configuration[12–14] stabilized by the strong DMI can be used to explain such a fast motion. Although some experimental efforts have been recently devoted to quantifing the DMI and the underlying physics, the mechanism of interfacial DMI is still elusive. Nevertheless, enhancement of DMI can be

---

[1)] Anni Cao and Xueying Zhang contributed equally to this work.

achieved by fine tuning of the interface configuration[15–17]. Therefore, effective control of DMI is essential in developing advanced storage class memory devices[18].

In this paper, we propose a method to control DMI by changing the thickness of the MgO layer in a Pt/Co/oxide system. The thickness of MgO of samples was tuned via a wedge structure and the corresponding effective DMI field was quantified by searching for its compensating in-plane field that minimized the DW motion velocity in creep regime. Samples with 0.2 nm Mg inserted between Co and MgO have also been prepared. We found that the variation of the effective DMI field can be attributed to the details of the Pt/Co interface, which indicates that the influence on adjacent interfaces of MgO could be used to regulate the interfacial DMI thanks to a large charge transfer and the interfacial electric field[19]. We also unveiled that the insertion of monoatomic Mg can help tuning of DMI through the MgO layer and get a large value up to 2.32 mJ/m$^2$.

**SAMPLE PREPARATION**

Samples with the structure of Ta(3 nm)/Pt(3 nm)/Co(1 nm)/MgO(t)/Pt(5 nm) and Ta(3 nm)/Pt(3 nm)/Co(1 nm)/Mg(0.2 nm)/MgO(t)/Pt(5 nm) were firstly deposited on a 500 μm Si wafer with a 300-nm thermal oxide layer by magnetron sputtering at room temperature, as shown in Figure 1. The (111) texture of the bottom Pt was ensured by a Ta seed layer[20], while the top Pt performed as a protective layer preventing the film oxidation. The base pressure of our ultrahigh vacuum deposition system is around $3 \times 10^{-8}$ mbar. In order to exclude the influence of variable growth conditions on the MgO quality, and thus to make the MgO thickness be the only variable in our system, the MgO layer was designed in a wedge structure. The thickness of MgO ($t_{\text{MgO}}$), which has been calibrated by an atomic force microscope (AFM), varies from 0.40 nm to 1.26 nm in 8 cm length for the samples without Mg layer, while $t_{\text{MgO}}$ varies from 0.20 nm to 2.00 nm in samples with a Mg insertion layer. In addition, a sample of the same structure without MgO layer was prepared as a reference. A sectional view by Cs-corrected Transmission Electron Microscopy and x-ray energy dispersive spectroscopy (EDS) curves are exhibited in Figure 1, (a) and (b) for samples without and with the ultrathin Mg layer. As shown in Figure 1 (a), except for the amorphous Co and MgO, each layer can be distinguished clearly. In contrast, Figure 1 (b) shows a much better resolution, as particularly clear in the inset which shows the Pt/Co/Mg/MgO/Pt part of the structure with a larger scale. The crystalline structure of the Pt/Co interface for samples with Mg is superior as compared to the other. Since the peaks of the EDS curves are indicative for the center of each layer, we can deduce that oxygen and magnesium permeated into the Co layer till some extend for both samples, but the insertion of a Mg layer separates cobalt and oxygen most effectively.

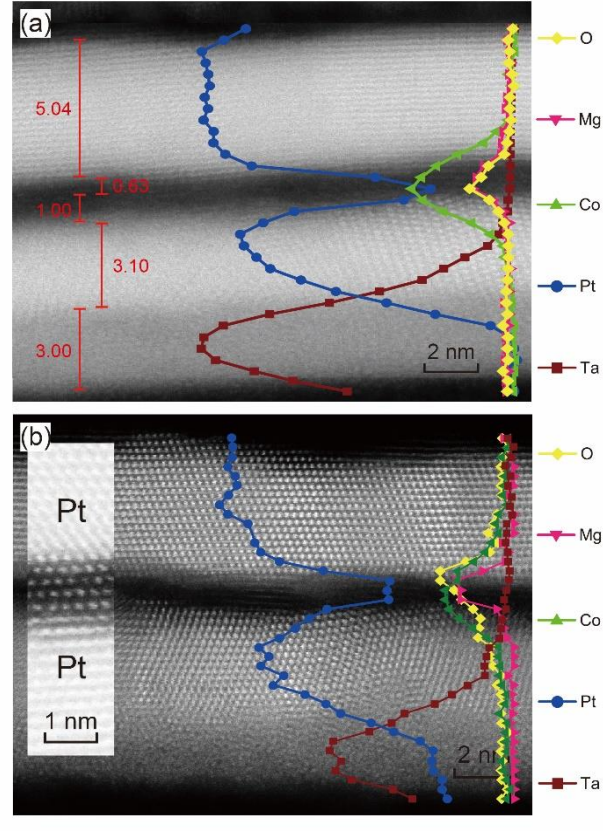

Figure 1. Cross profile of the samples (a) without Mg layer at $t_{MgO} \approx 0.60$ nm and (b) with Mg layer at $t_{MgO} \approx 0.80$ nm as measured by Transmission Electron Microscopy.

**MEASUREMENTS OF MAGNETIC PROPERTIES**

The wedge samples were cut into small squares with the lengths of the side equal to 2.5 mm for vibrating sample magnetometer (VSM) measurements. Figure 2 (a) and (b) show the hysteresis loops of two groups of samples with perpendicular magnetic fields at room temperature. It can be seen that all samples exhibit obvious perpendicular anisotropy. We can see that the saturation magnetization $M_S$ and the coercive field $H_C$ decrease with the increase of MgO thickness for the samples without Mg protection as shown in Figure 2 (c), while this decrease is avoided in samples with an inserted Mg layer (Figure 2(d)). The largest variation appears in the loops with the thinnest MgO and without MgO. The interfacial PMA can be calculated as $K_{eff} = \frac{1}{2}\mu_0 H_K M_S$, where $K_{eff}$ is the effective magnetic anisotropy energy, and $H_K$ is the effective anisotropy field[21,22] obtained by extracting the field corresponding to 90% of $M_S$ in curves with in-plane field. As shown in Figure 2 (c), the saturation magnetization $M_S$ decreases by 75% as the MgO thickness increases from 0 to 1.26 nm, which due to the partial oxidation of Co. Also, the proximity induced magnetic moment in the Pt will be quenched after inserting the MgO barrier. As shown in Figure 2(d), this significant shrinkage disappears when the Mg is inserted between Co and MgO. After insertion, the reduction of $M_S$ is only about 20%. The effective anisotropy field $H_K$ seems not to vary remarkable in both Figure 2 (c) and (d), so the trends of the effective magnetic anisotropy energy $K_{eff}$ are attributed to the variations of $M_S$ to some extent.

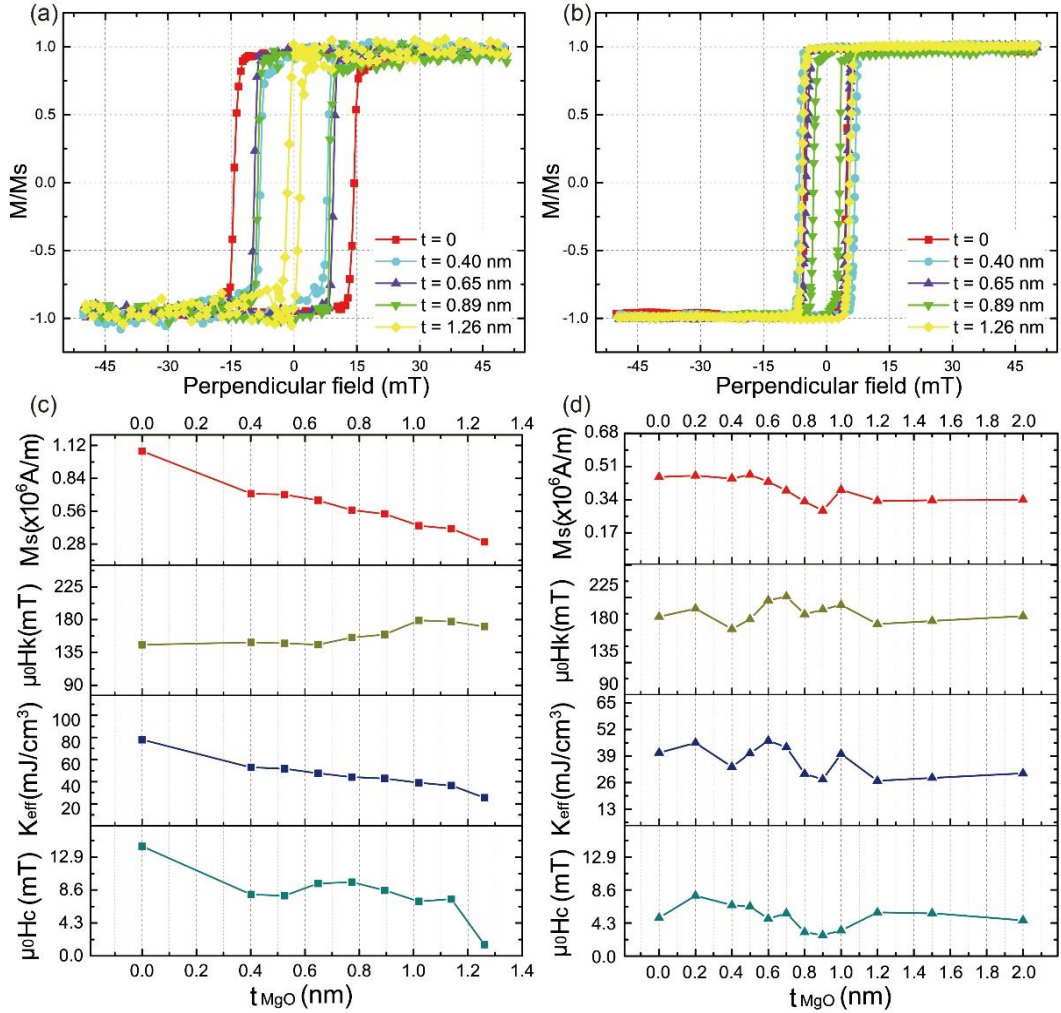

Figure 2. Hysteresis loops with perpendicular magnetic field for (a) the Ta/Pt/Co/MgO($t_{MgO}$)/Pt stacks and (b) the Ta/Pt/Co/Mg(0.2 nm)/MgO($t_{MgO}$)/Pt stacks with different MgO thicknesses. Magnetic properties obtained from the hysteresis loops for samples (c) without and (d) with Mg insertion layer.

**CHARACTERIZATION OF THE DMI**

In order to quantify the DMI in the various samples, we studied the asymmetric DW motion in the creep regime when an in-plane field was applied. In a sample with PMA and strong DMI, the magnetization in the center of DWs is fixed in a chiral direction by the effective DMI field $H_{DMI}$, i.e., DW configurations are of Néel type. The presence of an in-plane field will break this configuration, thus affecting the surface energy of DWs. This change will further influence the field driven DW velocity in the creep regime. For a fixed perpendicular driving field, the minimal DW velocity occurs when $H_{DMI}$ is completely compensated by the in-plane field. This compensating field can be seen as an indicator of the strength of DMI and was widely used to quantify $H_{DMI}$[23–25].

In our experiments, a Kerr microscope was used to measure the DW velocity. After saturating the sample, when an out of plane field $\mu_0 H_z$ was applied (~10 mT) in the opposite direction, domain bubbles will be nucleated and continuously expand. In creep mode, the DW expansion is not so fast ($10^{-4}$ ~ $10^{-5}$ m/s) so that we can capture Kerr images at a fixed time interval (0.08 s). Then the DW velocity can be calculated by analyzing this dynamic process. At the same time, an in-plane field $H_x$ could be applied and the DW velocity variation caused by $H_x$ could be probed, as shown in Figure 3. Since the perpendicular component of $H_x$ on the drive field $H_z$ would exponentially influence the $v$ (as shown in Fig. S1 S2 in the supplementary material), the sample was placed at the very center of the quadrupole magnet poles to avoid the probable interference to the measurement of velocity. In order to make sure that the applied in-plane field was exactly parallel to the sample and its stray field in the perpendicular was negligible, we have checked that when $H_z$ was removed, no DW motion would be seen even applied $\mu_0 H_x$ increased to a relatively high value (~300mT). Moreover, the leftmost velocity and rightmost velocity should exhibit symmetry about x-axis. Performing DW motion measurements this way, it was observed that as the thickness of MgO layer grows, more intensive DWs caused by interfacial defect appear, which indirectly hint the decrease of the coercive field $H_c$.

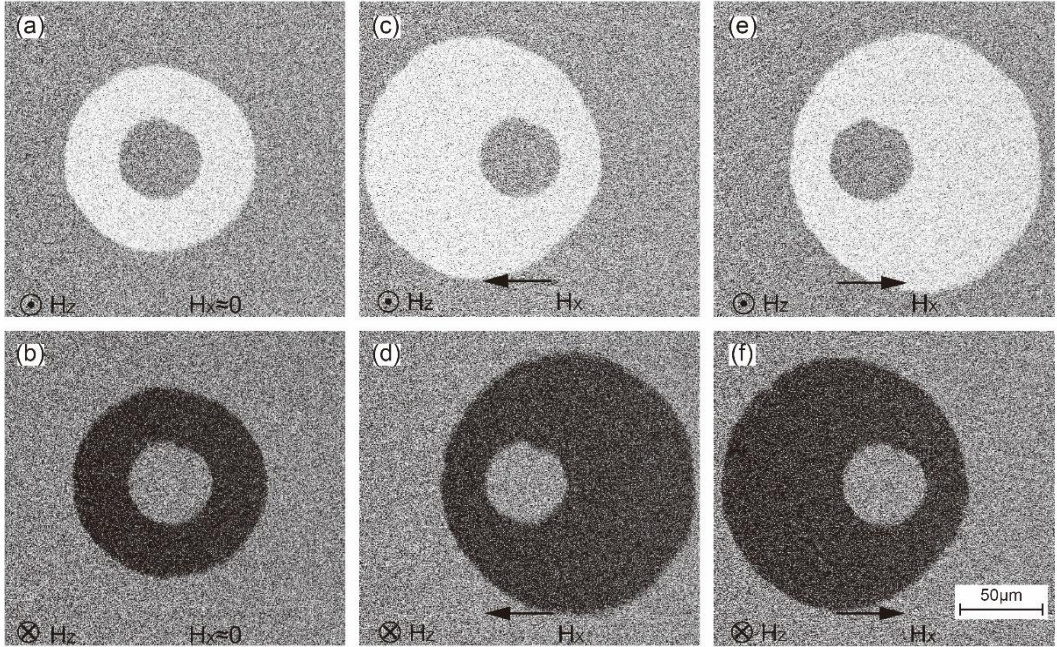

Figure 3. DW expansion of Pt/Co/Mg/MgO(0.6 nm)multilayers driven by out-of-plane magnetic field $\mu_0|H_Z| = 8$ mT. (a) (b) with in-plane field $\mu_0 H_x$ approximate to 0, (c) – (f) with $\mu_0|H_x| = 3270$ mT. Field directions have been marked in each image. Images in (a) – (f) were obtained by subtracting four images with specific time interval from a background with no DW. All images of DW were captured by a Kerr microscope.

A selection of our DW velocity measurements is shown in Figure 4. The dependence of the velocity on the in-plane field is found to be roughly quadratic, where the minimum occurs at a non-zero value of $H_x$. Reversing $H_z$, a very similar quadratic $H_x - v$ dependence if found, however, with a minimum occurring at the opposite value of $H_x$, as shown by black and red data points respectively. Similar $H_x - v$ measurements performed at different $H_z$ values are shown in Figure S3 in the supplementary material. We have estimated the influence of the inhomogeneity of the demagnetizing field and the stray field in different zones of the sample and found that this influence is negligible, see supplementary information. In order to exclude the direct influence of the thickness gradient on the DW motion velocity, we scaled the velocity at same rightmost place of the same domain and measured the same DW motion direction. Moreover, all measurements were focused on a zone of about 1.6×1.6 mm$^2$. In such a small zone, the variation of the MgO thickness is less than $2 \times 10^{-4}$ nm. So the velocity asymmetry caused by the structural gradient could be neglected. Therefore, we conclude that the velocity asymmetry shown in Figure 4 is mainly induced by DMI. Following the procedure in Ref. [30], $H_{DMI}$ is obtained from the in-plane field value corresponding to the lowest $v$, although we are aware of the problems of this method in certain cases[26,27].

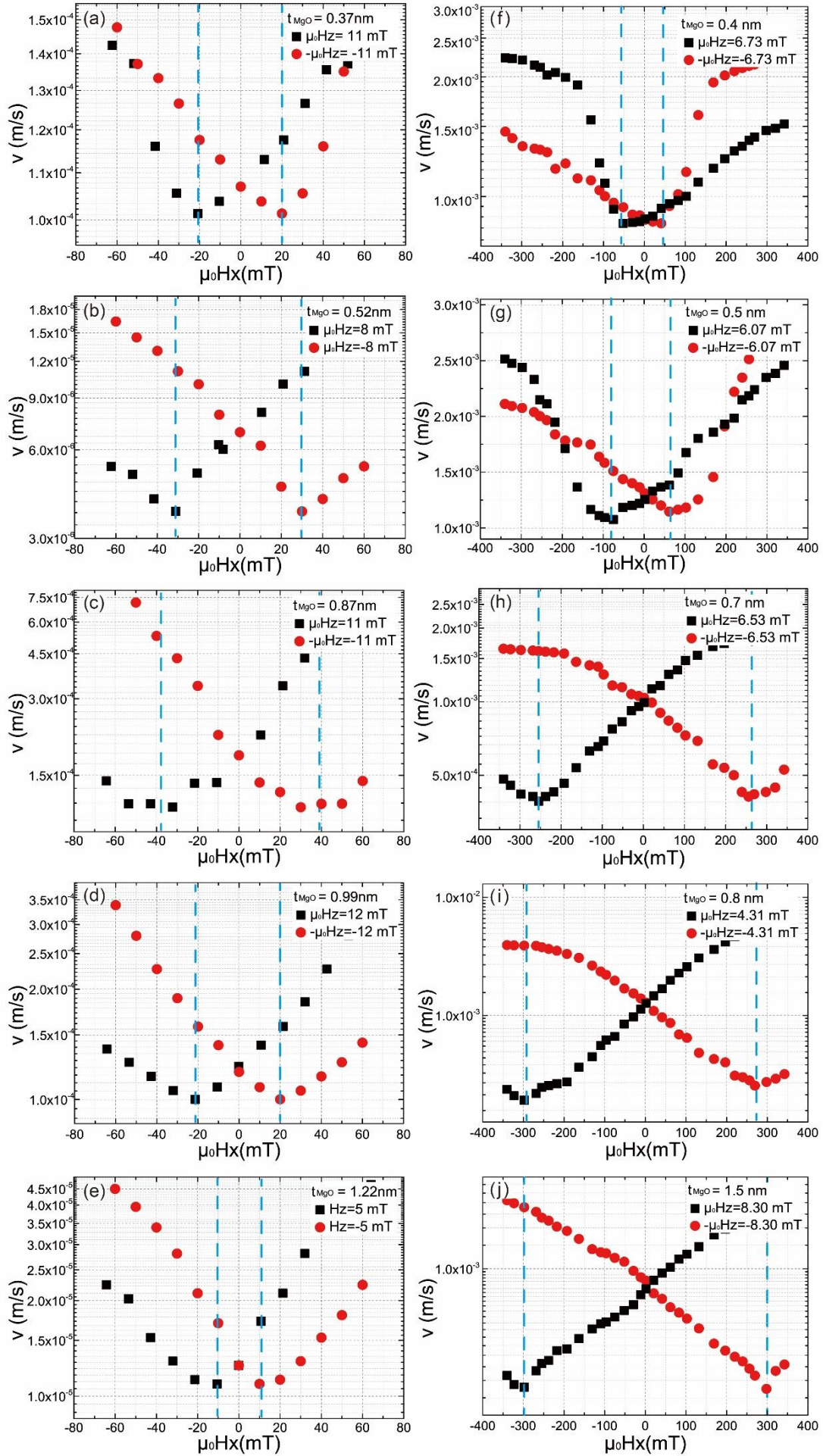

Figure 4. (a) – (e)Rightmost DW velocities of Pt/Co/MgO samples with $\mu_0 H_x$ varying from -60 to 60 mT, (f) – (j) Right-hand-edge DW velocities of Pt/Co/Mg/MgO samples with $\mu_0 H_x$ varying from -350 to 350 mT. Perpendicular fields ($\mu_0 H_z$) in both directions were measured. The blue vertical dashed lines stand for the place we got $\mu_0 H_{DMI}$.

The trends of the effective DMI field are depicted in Figure 5 with solid symbols. Since the $|H_{DMI}|$ of all samples with MgO layer in our experiments are stronger than the samples without MgO, the insertion of MgO enhances the DMI, especially for those samples with Mg inserted between Co and MgO. It is also found that, for the samples without Mg layer, the $H_{DMI}$ first increases as a function of MgO thickness, and after an optimum around 0.7 nm it decreases again. In contrast, $H_{DMI}$ for the samples with Mg layer grows and saturates at a relatively high level. By assuming an exchange stiffness constant $A = 15$ pJ/m[28,29] and substituting the experimental results into $|H_{DMI}| = |D|/(\mu_0 M_S \sqrt{A/K_{eff}})$[30], the absolute value of the DMI constant $|D|$ exhibits a similar trend as $|H_{DMI}|$ to both groups of samples. The maximum $|D|$ value of 0.77 mJ/m$^2$ occurs at the MgO thickness of 0.65 nm for samples without monatomic Mg. Surprisingly, the $|D|$ reaches up to 2.32 mJ/m$^2$ after the Mg inserted between the Co/MgO interface. The saturated $|D|$ value of samples with Mg is comparable to the published result of $D = 2.05$ mJ/m$^2$ in the same structure[31].

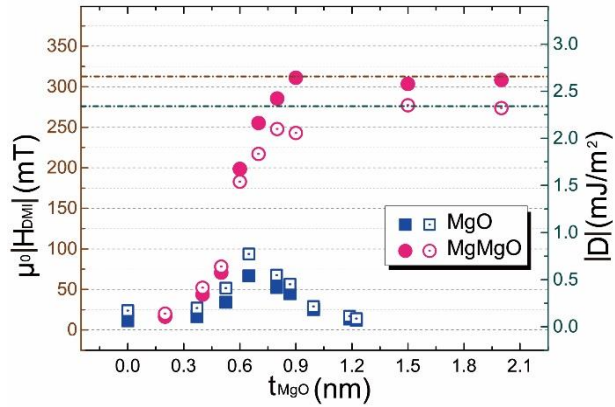

Figure 5. Trends of the effective DMI field and DMI constant as a function of $t_{MgO}$. Square symbols stand for Pt/Co/MgO($t_{MgO}$) samples while circular symbols stand for Pt/Co/Mg/MgO ($t_{MgO}$). Closed symbols stand for $\mu_0|H_{DMI}|$ and open symbols stand for $|D|$.

**DISCUSSION**

Using the above method, we carried out our measurements of $H_{DMI}$ versus MgO thickness on all samples with or without Mg layer; results were shown in Figure 5. For Pt/Co/MgO samples, the strengths of DMI exhibit first an increasing and then a decreasing trend, whereas a stable stage following an escalating trend can be seen for Pt/Co/Mg/MgO samples. In addition, the highest values of both $H_{DMI}$ and $|D|$ for samples with Mg are almost 3 times larger than the peak value for samples without Mg. However, here it is unexpected that when MgO thickness is larger than a critical thickness, the DMI starts to decrease. Comparing the experimental results of samples with Mg insertion layer, it can be inferred that excessive oxidation is the main reason for the DMI reduction. When MgO grows thicker, the stabilization of DMI for samples with Mg is more likely to reveal that the tuning of DMI is an interfacial effect.

As a primary dominant mechanism, upon inserting MgO, the structure is effectively changed from Pt/Co/Pt, which is nearly inversion symmetric and has a relatively weak DMI, towards Pt/Co/MgO, which is a prototype asymmetric structure with large DMI. As a consequence, the DMI shoots up after insertion of MgO. Secondly, further changes may be assigned to the interface between Co and Pt, and could involve several mechanisms. On the one hand, the oxidation of Co mentioned above decreases the effective Co thickness introducing DMI with the Pt layer[32]. On the other hand, the lattice mismatch between Co and MgO may induce strain effect to the Pt/Co interface[33,34] although the MgO layer is sputtered after the Co layer, MgO modifies the Pt/Co interface and influences the spin-orbit coupling (SOC) between the bottom Pt and Co. Last but the most important, as for the Co/MgO interface, according to density functional theory calculations, interfacial oxidation is related to a large charge transfer and to the large interfacial electric field that compensates the small spin-orbital coupling of the atoms at the interface, directly increasing the DMI[31,35]. It has been calculated that, different from the Pt/Co interface, the DMI and SOC energy source of the Co/MgO interface are localized in the interfacial Co layer[31,36], which indicates a diverse mechanism governed by the Rashba effect[37–39].

Noting that the influence of DMI on the free layer and the performance of the device has been intensively studied in recent years[40]. In addition, some devices based on the skyrmionic state in the free layer of a MTJ were proposed[41]. The TMR was also a critical factor when we detect skyrmions using an effective electrical method[42,43]. As, the HM/FM/MgO structure we studied is very similar to the configuration of the tunnel barrier layer / free layer / capping layer of the most popular MTJ structure[44], our study could be therefore very useful to investigate the properties and effects of DMI for the electrical nucleation and detection of magnetic skyrmions through MTJ. Not only Co, but also the influence of MgO on adjacent interfaces could be used to fine-tune the DMI in Pt/Co/MgO samples which is valuable for the induction of chiral magnetic order.

**CONCLUSION**

In conclusion, varying the thickness of MgO in a Pt/Co/MgO material system can significantly affect the Dzyaloshinskii−Moriya interaction. The DW motion was measured to characterize the strength of DMI, which is caused by the structural asymmetry between the top interface and bottom interface. We found that indeed it is possible to tune DMI by varying the MgO thickness, especially when Co is protected by an ultrathin Mg insertion layer. Not only the Co/MgO interface but also the Pt/Co interfaces were changed. This study will be very helpful for thin film design to obtain large DMI and stabilize the magnetic skyrmions with expected size for memory and logic devices.

**ASSOCIATED CONTENT**

Additional experiment results of DW motion, calculation of demagnetizing field and stray field were seen as the **Supporting Information**.

**AUTHOR INFORMATION**

**Corresponding Author**
*E-mail: weisheng.zhao@buaa.edu.cn (W.-S.Z.).

**Acknowledgment.** The authors would like to thank the supports by the projects from National Natural Science Foundation of China (No. 61571023, 61501013 and 61471015), Beijing Municipal of Science and Technology (No. D15110300320000) and the International Collaboration Project from the Ministry of Science and Technology in China (No. 2015DFE12880), the Program of Introducing Talents of Discipline to Universities in China (No. B16001).

# Supporting Information for:
# Tuning the Dzyaloshinskii−Moriya Interaction in Pt/Co/MgO heterostructures through MgO thickness

Anni Cao,[a,1)] Xueying Zhang,[ab,1)] Bert Koopmans,[c] Shouzhong Peng,[a] Yu Zhang,[ab] Zilu Wang,[a] Shaohua Yan,[a] Hongxin Yang,[d] and Weisheng Zhao[*a]

[*] Corresponding authors

[a] Fert Beijing Institute, BDBC, School of Electronic and Information Engineering, Beihang University, Beijing, China

[b] Centre for Nanoscience and Nanotechnology, University Paris-Saclay, Orsay, France

[c] Department of Applied Physics, Institute for Photonic Integration, Eindhoven University of Technology, Eindhoven, The Netherlands

[d] Key Laboratory of Magnetic Materials and Devices, Ningbo Institute of Materials Technology and Engineering, Chinese Academy of Sciences, Ningbo, Zhejiang, China

## I. Exponentially-varied $v$ versus $H_z^{-\frac{1}{4}}$

The determination of DMI by the asymmetry of DW velocity is based on the creep mode DW motion. We measured the DW velocity $v$ with different driven field value $H_z$ to improve the feasibility of the measurement, according to the characterization of the creep mode motion which could be described as

$$v \sim exp\left(-\alpha H_z^{-\frac{1}{4}}\right) \tag{S1}.$$

The results of samples without and with Mg insertion layer were shown below. All the driven fields we use to measure the $H_{DMI}$ circled by red-dotted line are inside the linear regions in Figure S1 S2.

---

[1)] Anni Cao and Xueying Zhang contributed equally to this work.

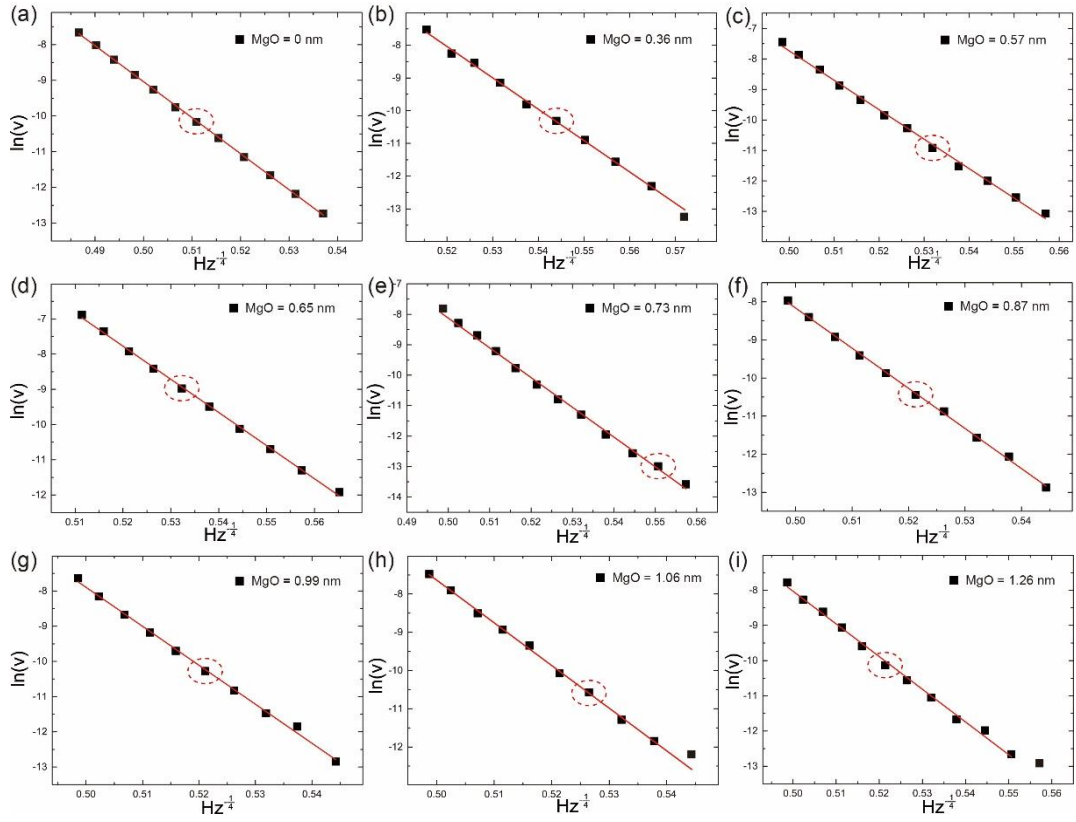

Figure S1. Fitting lines of $H_z^{-\frac{1}{4}} - \ln(v)$ for samples with different MgO thickness when there is no Mg inserted.

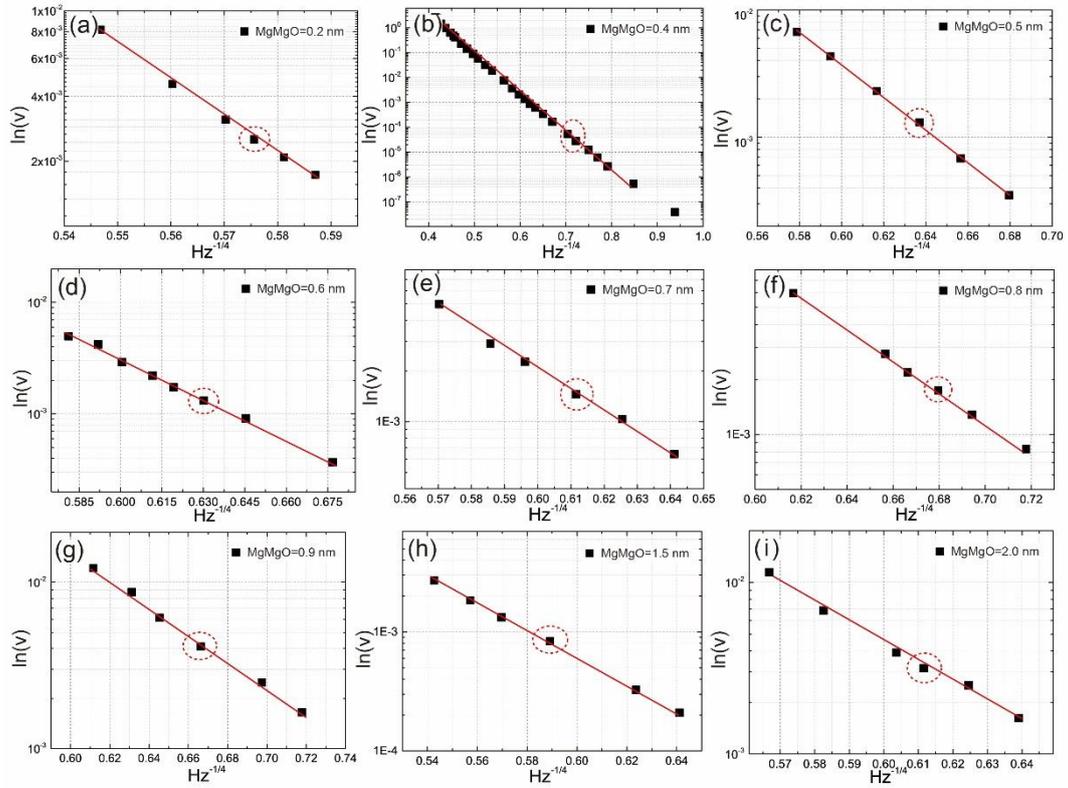

Figure S2. Fitting lines of $H_z^{-\frac{1}{4}} - \ln(v)$ for samples with different MgO thickness when there is 0.2 nm Mg inserted.

## II. $H_{DMI}$ with different $H_z$

In order to prove that the driven field $H_z$ has no effect on the effective DMI field $H_{DMI}$, the equi-speed contour maps of $t_{MgO} = 0.40, 0.52$ and $0.77\ nm$ samples with 1.00 nm Co layer were shown in the Figure S6.

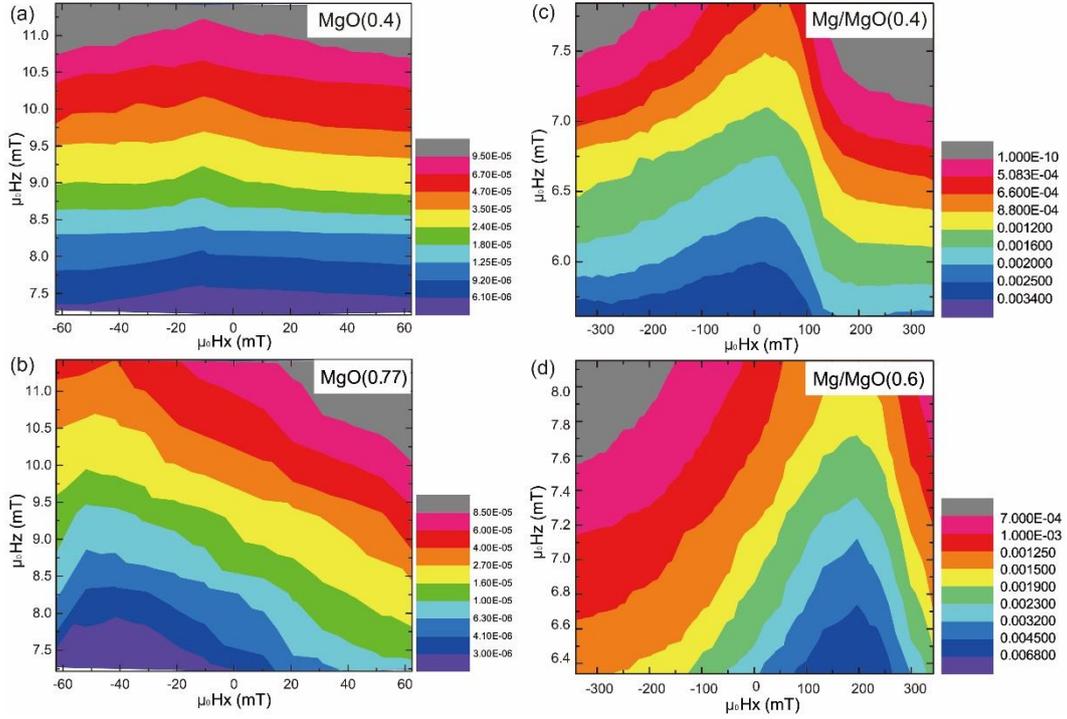

Figure S3. Two-dimensional equi-speed contour map of $v$ as a function of $\mu_0 H_x$ and $\mu_0 H_z$. The color corresponds to the magnitude of $v$ with the scale on the right.

## III. Calculation of the demagnetizing field and the stray field

Using the concept of magnetization current[1,2], we numerically calculated the demagnetizing field as shown in Figure S7 (a), magnetic domains can be infinitely divided into small magnetic elements. Each element is equivalent to a ring current element, with current per unit area $I = t * M_S$, where $t$ is the thickness of magnetic layer and $M_S$ is the saturated magnetization per volume. For a uniformly magnetized out-of-plane domain, the current of one magnetic element can always be cancelled out by the current of the neighboring ones, except for elements in the boundary of domains and the edge of the sample. Namely, the magnetization currents are zero inside or outside the domain. The only non-zero contributions are on the boundary of the domain and the edge of the sample which are together noted as demagnetizing field here.

The contribution of the domain boundary is equal to the Oersted field produced by the effective current at the edge of domains. For a magnetic bubble with radius $r$ in a magnetic thin film, the electrical circuit is plotted in Figure S4 (b). In addition, the DW width in PMA samples[3] could be estimated as 10 nm, and we have testified the width wouldn't strongly influence the demagnetizing field. This width is considered as the distance between two domains with opposite magnetization, i.e. the distance between the two circular current circuits. Similarly, the contribution of the sample edge is closely related to the distance $d$ between DWs and the edge of sample, as showed in Figure S4 (b).

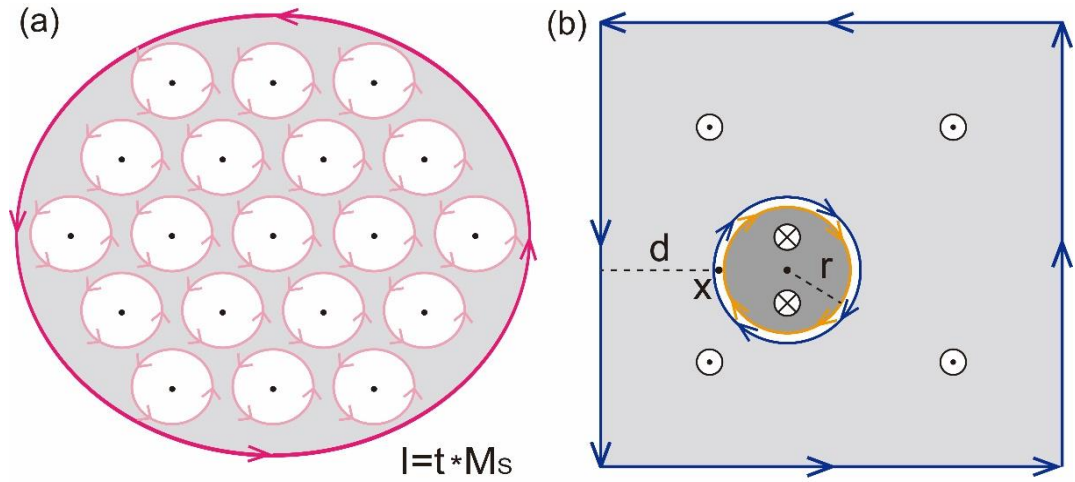

Figure S4. Magnetization current of the magnetic structure. (a) Sketch to show the concept of magnetization current. (b) Electrical circuit identical to the current created by the magnetization at the edges of the structure and along the DW. Then, the stray field can be calculated using the Biot-Savart's law.

Using MATLAB, we respectively calculated the demagnetizing field $\mu_0 H_{de} = \mu_0 H_{DW} + \mu_0 H_{sample}$ at point X shown in Figure S4, with $r$ varying from 100 μm to 250 μm (the size range of DW we observed) and $d$ varying from 120 μm (the minimum distance of our photo results) to $5 \times 10^6$ μm (the measured sample size). The results are plotted in Figure S5. It can be seen that, with the growth of bubble domains the demagnetizing field $\mu_0 H_{de}$ is lower than 0.023 mT, and as the distance to sample edge increases, the variation of $\mu_0 H_{de}$ is always less than 0.002 mT which can be neglected.

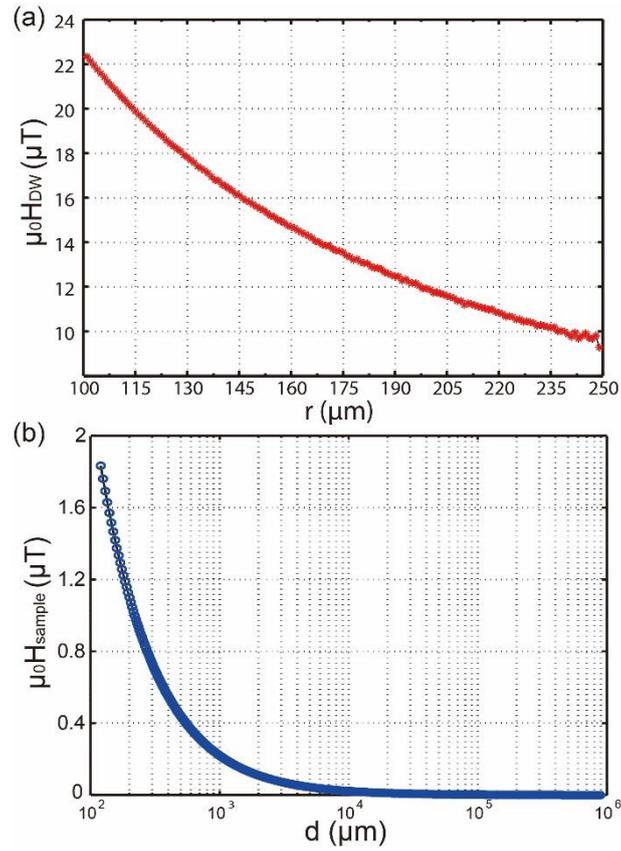

Figure S5. Numerically calculated demagnetizing field at point P as a function of the inverse of bubble radius, (a) the Oersted field of the effective current at the boundary of the observed DW $\mu_0 H_{DW}$, (b) the Oersted field of the effective current at the sample edge $\mu_0 H_{sample}$.